# THERMAL CHARACTERIZATION OF ABSORBING COATINGS FOR THERMAL DETECTORS OF RADIATION BY PHOTOPYROELECTRIC METHOD

Svetlana L. Bravina, Nicholas V. Morozovsky, Galina I. Dovbeshko
and Elena D. Obraztsova[a]

Institute of Physics NASU, 46 Prospect Nauki, Kiev, 03028, Ukraine
bravmorozo@yahoo.com

[a]Natural Science Centre of A. M. Prokhorov, General Physics Institute RAS, 38 Vavilov str.,
Moscow, 119991, Russia

## ABSTRACT

By photothermomodulatoin method the comparative study of thermal diffusivity of absorbing coating for sensitive elements of pyroelectric detectors of radiation formed from metal dispersion layer blacks, dielectric paint blacks and carbon nanotubes paint blacks has been performed. Prospects of using carbon nanotubes based black absorbing coatings for pyroelectric and other thermal detector application are shown.

## 1. INTRODUCTION

Application field of one of the most important class of thermal detectors, pyroelectric detectors of radiation (PDR), with improving their manufacture technology includes all modern engineering branches from space and defense equipment to household appliances, public and individual security and fire alarm systems[1-9].

The principle of operation of pyroelectric detectors of radiation is based on the temperature dependence of polarization of sensitive element (SE) material and registration of electrical charge connected with polarization change induced by temperature variation[2, 3, 5, 7-9]. Radiation absorption by the surface or volume of SE is the first and very important stage of radiation conversion into the electrical response of detectors[2, 3, 5]. That is why sensitive and frequency characteristics of commonly used PDR with surface absorption are in a high degree determined by thermophysical parameters of absorbing coating (AC) of SE[10-12].

Well-known technologies for making up the AC of PDR **use** precious (Pt, Au and Ag) and non-precious (Ni, Cr and Ni-Cr) metals evaporated in the conditions of "imperfect" vacuum and deposited in certain atmosphere and temperature conditions[10, 11] and also different lacquer paints[12] formed by smearing or pulverization. These so called metal and dielectric blacks possess sufficient spectral characteristics in the near- and middle- IR range[10, 12]. However, these blacks have the relatively low thermal conductivity and restrict the speed of response of PDR by the time of thermal diffusion through the layer of AC[11].

Earlier dense enough and so low-resistance golden black was used as a absorbing electrode[13]. Progress in modern technology allows enriching AC stock with composite materials based on carbon nanotubes (CNT)[14] which are mainly characterized by the combination of high values of thermal and electrical conductivity[15]. The progress of corresponding methods of deposition of such types of AC can result in the development of absorbing CNT-electrodes that combine low electrical and thermal resistance for the thickness value enough for absorption in middle- and far- IR-range.

Earlier we presented the results of the comprehensive investigation of AC based on thin layers of SiO<Cu> metal-oxide composite and incorporated small particles of metals, obtained by methods of coevaporation[15].

In this paper we present the results of the comparative study of thin layer of absorbing coatings based on metal dispersive layer and dielectric paint blacks and also carbon nanotubes paint black as promising for using for sensitive elements pyroelectric detectors and for other types of thermal detectors of radiation.

## 2. EXPERIMENTAL

### 2.1. Sample preparation

Metal black AC were manufactured by evaporation of corresponding metal (gold in this experiment) and deposition of its vapors on the SE electrode surface under imperfect vacuum condition of 0,1 - 10 Torr in a nitrogen atmosphere with a low content of oxygen at the substrate temperature of about 200 K[16].

For the manufacture of dielectric lacquer-paint (LP) black AC was used the solution prepared from industrial multipigment black lacquer-paint (LP) coating dried preliminary.

For the manufacture of CNT based black AC the carbon single wall nanotubes (SWNT) prepared by method of arc discharge in the He-atmosphere[14] were used. Then suspension on the base of SWNT crumbled preliminary, organic binder and dissolvent was prepared.

The LP-black AC and CNT-black AC were prepared by depositing thin layer of solution of a low viscosity on the SE electrode surface and subsequent evaporating solvent at heightened temperature. After drying on the electrode surface were obtained the layers of 5-20 μm thickness close to that of Au-black AC. For manufactured CNT-black AC is characteristic the value of 20-50 kΩ/□ surface resistance and a high degree of blackness close to that of LP-black and Au-black AC.

For pyroelectric investigations were formed SE with AC deposed directly on the main surfaces of $LiNbO_3$ plates of polar Z-cut with evaporated Cu/Cr-electrodes of 20-50 mm$^2$ area. Such SE with various types of AC were connected by means of a ring-shape holder to the input of the FET matching stage[17] with variable from high (~10 GΩ at the frequency of 20 Hz) to relatively low (~100 kΩ) input impedance value.

## 2.2. Measurements

The pyroelectric response measurements were performed by the modulation photopyroelectric thermowave method[18] in the frequency range 10 Hz ≤ $f_m$ ≤ 100 kHz of modulation of IR radiation flux. The measuring system for thermowave probing allowed us to obtain amplitude-to-frequency $U_\pi(f_m)$ and phase-to-frequency $\varphi_\pi(f_m)$ dependences of pyroelectric response in two operation modes of pyroelectric current and pyroelectric voltage characteristic for PDR operation. Due to the fact that in pyroelectric current mode $U_\pi = U_{\pi1} \propto \gamma/c_1$ and in pyroelectric voltage mode $U_\pi = U_{\pi2} \propto \gamma/c_1\varepsilon f_m$, where $\gamma$ is the pyroelectric coefficient, $c_1$ is the volume heat capacity, there is a possibility to evaluate the dielectric permittivity $\varepsilon$ from pyroelectric measurements[2, 16-18] by introducing the dielectric ratio $D_\pi = U_{\pi1}/U_{\pi2}f_m \propto \varepsilon_\pi$.

The connection of amplitude and phase of pyroelectric response with the thermal parameters of both SE of PDR and AC material in consequence of fundamental frequency dependence of the length of temperature wave $\lambda_T = (a_T/\pi \cdot f_m)^{1/2}$, where $a_T$ is the thermal diffusivity[16, 18] allows one to estimate the value of $a_T$ for material of absorbing coating by analysis of $U_\pi(f_m)$ and $\varphi_\pi(f_m)$ dependences.

### 3. RESULTS AND DISCUSSION

In Figure 1 we presented the dependences of $U_{\pi 1, 2}(f_m)$ and $\varphi_{\pi 1, 2}(f_m)$ for SE of PDR with one-side AC from Au-black.

Significant difference in $U_{\pi 1,2}$ levels for the electrodes with AC and without it at the low frequencies corresponds to a good quality of Au-black.

Flat frequency dependences of $U_{\pi 1}$, $U_{\pi 2} \cdot f_m$, $D_\pi$ and $\varphi_{\pi 1,2}$ under irradiation from the side of Cu-electrode (Fig. 1, left side) correspond to the thermal, polar and dielectric uniformity. In this case the following dependences are observed[5, 15, 17]:

$U_\pi = U_{\pi 1} \propto \gamma/c$ and $U_{\pi 1}(f_m) = \text{const} \, (f_m)$ in the pyroelectric current mode; and

$U_\pi = U_{\pi 2} \propto (\gamma/c\varepsilon)/f_m$ and $U_{\pi 2}(f_m) \cdot f_m = \text{const} \, (f_m)$ in the pyroelectric voltage mode, as $U_{\pi 1}/U_{\pi 2} \cdot f_m = \text{const} \, (f_m)$, so $D_\pi(f_m) = U_{\pi 1}(f_m)/U_{\pi 2}(f_m) \cdot f_m \propto \varepsilon_\pi = \text{const} \, (f_m)$ ,

and also $\varphi_{\pi 1,2}(f_m) = \text{const} \, (f_m)$ and $\varphi_{\pi 2} - \varphi_{\pi 1} = 90^\circ$.

The noticeable approach of $\varphi_{\pi 1}(f_m)$ to $\varphi_{\pi 2}(f_m)$ and $U_{\pi 1}(f_m)$ to $U_{\pi 1}(f_m)$ under $f_m$ increasing is connected with deviation from the conditions of pyroelectric current mode. Under that with $f_m$ increasing the SE impedance becomes less than the load resistance $R_l$ and the inequality $2\pi f_m R_l C \ll 1$ necessary for performing pyroelectric current mode changes for the opposite $2\pi f_m R_l C \gg 1$, characteristic for pyroelectric voltage mode, that leads to the coincidence of the dependences.

The weak high–frequency drop of $U_{\pi 2}(f_m) \cdot (f_m)$, $U_{\pi 1}(f_m)$ and increase of $\varphi_{\pi 2}(f_m)$ noticeable at the high–frequency interval corresponds to the existence of under-electrode non-homogeneity. Taking into consideration the known for $LiNbO_3$ value of $a_T = 1{,}5 \cdot 10^{-6} \, \text{m}^2/\text{s}$[5], with regard to the value of the $\lambda_T$ for $f_m = 40$ kHz, is possible to evaluate the thickness of this non-homogeneity as about 3 μm

Under irradiation from the side of Au-black (Fig. 1, right side) the flat frequency dependences of $U_{\pi 1}$, $U_{\pi 2} \cdot f_m$ and $\varphi_{\pi 1,2}$ are characteristic at low frequencies which corresponds to inequality $\lambda_\tau/L \gg 1$ between the length of temperature wave and the thickness L of the AC layer. At frequencies higher than 1 kHz we observed a significant decrease of response amplitude and additional positive phase shift. Such peculiarities are connected with a thermal damping action of AC, which is manifested under approaching $\lambda_\tau$ to L with increase of $f_m$ value. Under further increase of $f_m$ the inequality $\lambda_\tau/L < 1$ is getting valid. Strengthening of this inequality leads to the decrease of exponential in respect to $f_m^{1/2}$ multiplier for response amplitude $\Delta U_\pi(f_m) \propto \exp(-L/\lambda_\tau) = \exp(-L(\pi f_m/a_\tau)^{1/2})$ and increase of a linear in respect to $f_m^{1/2}$ contribution to the phase $\Delta\varphi_\pi(f_m) = L/\lambda_\tau = L(\pi f_m/a_\tau)^{1/2}$ As a result with increasing $f_m$ the level of pyroelectric response under irradiation from the Au-black side becomes even lower than under irradiation from the side of the Cu-electrode (see Fig. 1, left and right sides).

In Figure 2 we presented $U_{\pi 1, 2}(f_m)$ and $\varphi_{\pi 1, 2}(f_m)$ dependences for SE of PDR with AC from LP-black and AC from CNT-black.

Under irradiation of the side with LP-black AC (Fig. 4, left side) the behavior of $U_{\pi 1,2}(f_m)$ and $\varphi_{\pi 1,2}(f_m)$ dependences is similar to that observed for Au-black AC, namely the region of high-frequency drop of amplitude-frequency dependence is observed for $f_m > 1$ kHz.

Under irradiation of the side with CNT-black the dependences of $U_{\pi 1, 2}(f_m)$ and $\varphi_{\pi 1, 2}(f_m)$ are less sharp than observed in the case of irradiation as Au-black AC as LP-black AC (see Fig. 1 and Fig. 2). Near the same levels of $U_{\pi 1, 2}$ for Au-black AC, LP-black AC and CNT-black AC at low frequencies corresponds to good absorption of CNT-black AC.

The high–frequency drop of $U_{\pi 2}(f_m)\cdot(f_m)$ and increase of $\varphi_{\pi 2}(f_m)$ for CNT-black AC connected with thermal damping are not so sharp and are observed at the frequencies of 1 orders higher than those ones for AC from Au-black (compare Fig. 1 and Fig. 2).

Obtained $U_{\pi 1, 2}(f_m)$ and $\varphi_{\pi 1, 2}(f_m)$ dependences for SE with different AC allows one to estimate the value of thermal diffusivity of each AC material. It can be done by using the characteristic dependences of $U_\pi(f_m)$ and $\varphi_\pi(f_m)$ in the frequency range where the thermal damping is significant. For Au-black AC , LP-black AC and for CNT-black AC these estimations are presented in the Table 1.

Obtained $a_T$ values for CNT-paint black are at least of 1 order of value higher than $a_T$ values for other black AC under investigation and even are higher than $a_T$ values for massive graphite.

It can be connected with the high contribution of relatively large regions of high contacted CNT-bundles interconnected through developed surrounding of small CNT-bunches.

Combination of enhanced thermal parameters in comparison with those for known AC together with foreknew of its further progress and of similar spectral characteristics near to those for Au-black allows us to consider the carbon nano-tube based black as promising for absorbing coatings of thermal detectors of radiation.

TABLE 1. Thermal Diffusivities of Black Absorption Coatings for Pyroelectric Detectors, Dielectric and Resistive Bolometers

| Type of theCoating | Thickness | Thermal Diffusivity |
|---|---|---|
| Au-black Dispersion layer | 2 – 10 μm | $(1 - 3).10^{-7}$ m²/s |
| Dielectric Paint Black | 3 – 20 μm | $(2 - 4).10^{-7}$ m²/s |
| Carbon Nano-Tube Paint Black | 5 – 20 μm | $(3 - 6) \, 10^{-6}$ m²/s |
| Graphite (extruded) | 0,2-0,5 mm | $(2 - 7) \, 10^{-5}$ m²/s |
| **For comparison** | | |
| **Material** | **State** | **Thermal Diffusivity** |
| Graphite | Plates | $\parallel 1,3 \cdot 10^{-3}$ m²/s  $\perp 4 \cdot 10^{-6}$ m²/s |
| Diamond | Single crystal | $1,1 \, 10^{-3}$ m²/s |

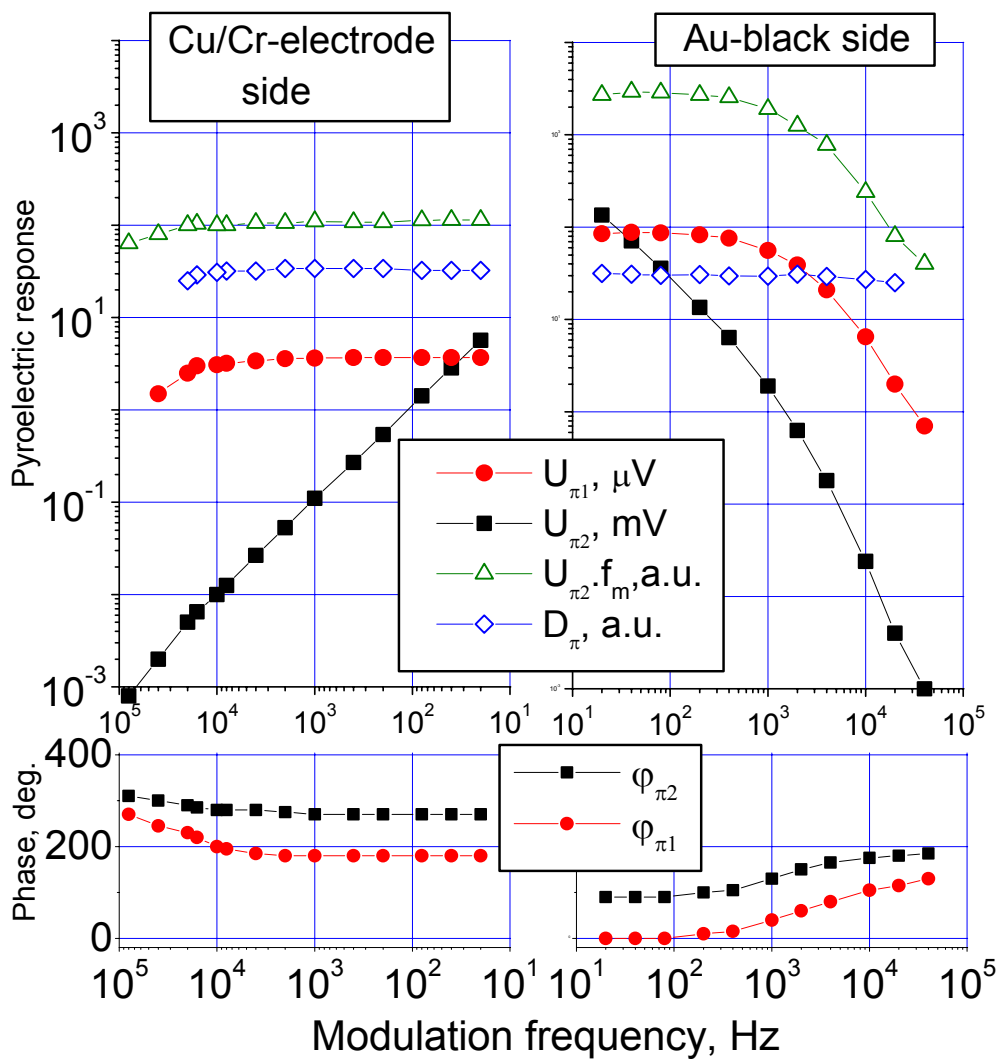

Fig. 1. Dependences of $U_{\pi 1}(f_m)$, $U_{\pi 2}(f_m)$, $U_{\pi 2}(f_m) \cdot f_m$ and $U_{\pi 1}(f_m)/U_{\pi 2}(f_m)$ $f_m$ and also $\varphi_{\pi 1}(f_m)$ and $\varphi_{\pi 2}(f_m)$ for SE of PDR with one-sided AC from Au-black. (Cu/Cr-electroded, 200 μm LiNbO$_3$ sheet)

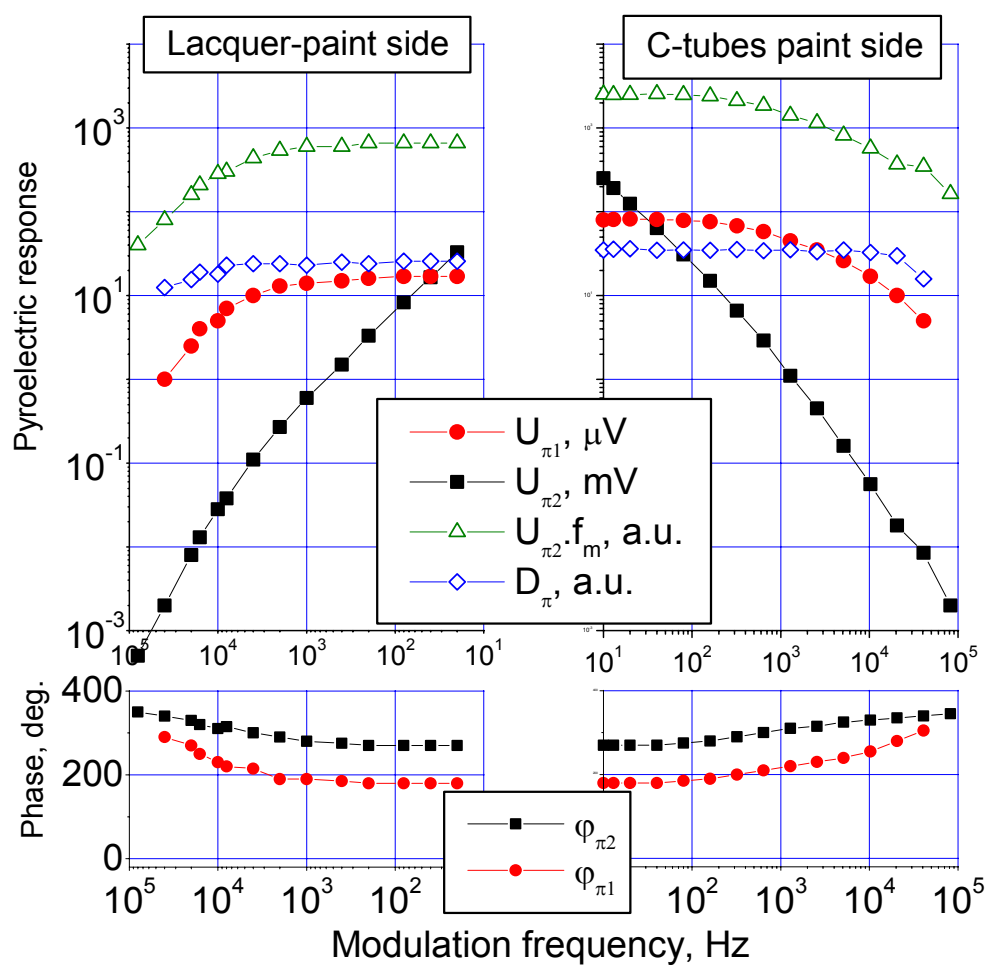

Fig. 2. Dependences of $U_{\pi1}(f_m)$, $U_{\pi2}(f_m)$, $U_{\pi2}(f_m)\cdot f_m$ and $U_{\pi1}(f_m)/U_{\pi2}(f_m)$ $f_m$ and also $\varphi_{\pi1}(f_m)$ and $\varphi_{\pi2}(f_m)$ for SE of PDR with AC from lacquer paint black and AC from CNT-paint black. (200 μm sheets; left: LiNbO$_3$, right: LiTaO$_3$)